\documentclass[%
 reprint           , 
 superscriptaddress,
 altaffilletter    ,
 nofootinbib,
 aps               ,
]{revtex4-1}

\usepackage{graphicx}
 \graphicspath{{./}}

\usepackage{multirow}
\usepackage{dcolumn} 

\usepackage{amsmath}
\usepackage{amssymb}
\usepackage[version=3]{mhchem}
\usepackage{siunitx}

\usepackage[hidelinks,linktocpage,colorlinks]{hyperref}
\hypersetup{
            colorlinks,
            linkcolor={red!70!black}  ,
            citecolor={green!70!black},
            urlcolor ={blue!70!black}
           }

\usepackage{comment} 
\usepackage[switch]{lineno} 
\usepackage[dvipsnames]{xcolor}
\usepackage{xspace}

\newcommand{\al}    {\ensuremath{\alpha}\xspace}
\newcommand{\bt}    {\ensuremath{\beta}\xspace}
\newcommand{\gm}    {\ensuremath{\gamma}\xspace}

\newcommand{\YAG}  {\ce{YAG{:}Ce}\xspace}
\newcommand{\Am}    {\ce{^{241}Am}\xspace}
\newcommand{\Na}    {\ce{^{22}Na}\xspace}

\newcommand{\QF}   {QF\xspace}

\newcommand{\uniMIB} {\affiliation{Department of Physics, University of Milano-Bicocca, Milano I-20126, Italy}}
\newcommand{\infnMIB}{\affiliation{INFN -- Sezione di Milano-Bicocca, Milano I-20126, Italy}}

\begin{document}

 \title{Comprehensive characterization of a YAG:Ce scintillator: light yield, alpha quenching and pulse-shape discrimination}

 \author{L.\ Gironi}   \uniMIB \infnMIB
 \author{S.\ Dell'Oro}\email{stefano.delloro@mib.infn.it} \infnMIB
 \author{E.\ Giussani} \uniMIB \infnMIB
 \author{C.\ Gotti}            \infnMIB
 \author{E.\ Mazzola}  \uniMIB \infnMIB
 \author{M.\ Nastasi}  \uniMIB \infnMIB
 \author{D.\ Peracchi} \uniMIB \infnMIB


 \keywords{}

\begin{abstract}
 Solid-state scintillators are widely used in particle and applied physics due to their versatility and resistance to diverse environments and operating conditions.
 This broad range of applications calls for thorough characterization of scintillating crystals.
 Among these materials, cerium-doped yttrium aluminum garnet (\YAG) is a promising scintillator owing to its favorable timing characteristics, high light yield, good mechanical properties, and chemical stability.
 In this work, we report a comprehensive experimental characterization of a \YAG crystal exposed to both \gm and \al radiation.
 We extract the scintillation decay time and light yield, and study their evolution from room temperature down to approximately $-50 ^\circ$ C.
 We perform a detailed investigation of the quenching factor for \al particles in the energy range from about $6$ MeV down to $1$ MeV, finding a value that decreases from approximately $0.17$ to $0.10$.
 We also explore the possibility of pulse-shape discrimination based on the different signal evolution depending on the interaction type, demonstrating strong classification capabilities.
 These results provide a detailed assessment of the performance of \YAG for radiation-detection applications and offer insight into its potential use in environments requiring reliable particle identification and stable response across a wide range of operating conditions.
 \\[+9pt]
 Published on: J.\ Instrum. {\bf 21}, P06017 (2026) \hfill DOI: \href{https://doi.org/10.1088/1748-0221/21/06/P06017}{10.1088/1748-0221/21/06/P06017}
\end{abstract}

\maketitle

\section{Introduction}

 Solid-state scintillating crystals remain among the most versatile and reliable detectors for the measurement of ionizing radiation.
 Their combination of high light yield, mechanical robustness, and compatibility with compact photodetectors makes them indispensable in applications ranging from medical diagnostics to high-energy and astroparticle physics.

 Continuous progress in crystal growth techniques and dopant engineering have led to scintillators with improved energy resolution, faster decay times, and enhanced radiation hardness.
 At the same time, the introduction of novel materials and optimized compositions calls for systematic and quantitative approaches to assess their suitability for precision spectroscopy and timing applications.
 A comprehensive characterization of scintillating crystals must therefore include measurements of their light output, emission spectrum, decay kinetics, and response linearity to radiation deposition.

 The dependence of the scintillation yield on the ionization density of the incident radiation plays a particularly important role. It is well established that highly-ionizing particles, such as \al particles or heavy ions, produce less scintillation light per unit deposited energy than electrons or \gm rays~\cite{Birks:1951boa}.
 This phenomenon, commonly referred to as scintillation \emph{quenching}, arises from the increased probability of non-radiative recombination processes along dense ionization tracks.
 Quenching plays a central role in both detector calibration and particle identification.
 On the one hand, it introduces non-linearities that must be accurately quantified and corrected in order to achieve reliable energy reconstruction.
 On the other, it can be exploited to discriminate among different classes of radiation through pulse-shape analysis.
 A precise determination of this effect is therefore essential for low-background and rare-event experiments.

 \begin{figure*}[tb]
  \includegraphics[width=1.\textwidth]{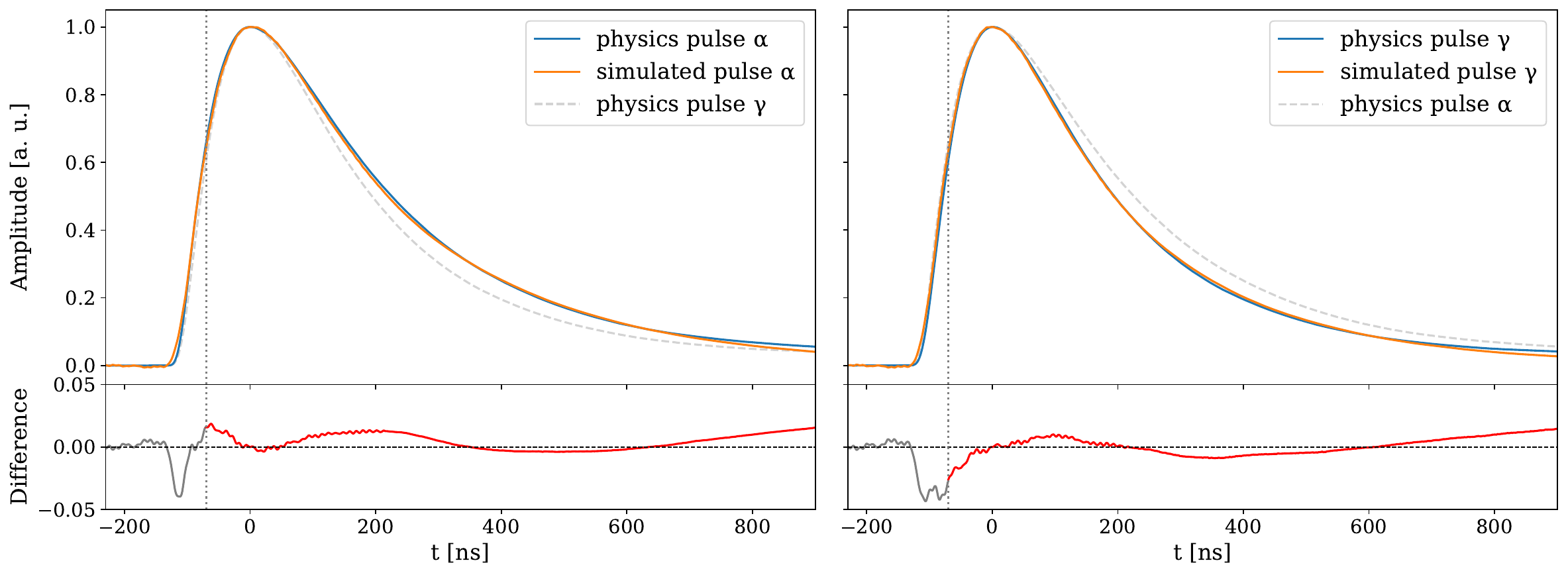}
  \caption{Comparison of normalized experimental pulses (blue) and pulses constructed from single-cell templates (orange); the absolute difference between the pulses is shown in the bottom panels.
  {\it (Left)} Average pulse from \al events corresponding to the $5.49$-MeV peak of \Am.
  {\it (Right)} Average pulse from \gm events corresponding to the $511$-keV peak of \Na.
  In both panels, the reference pulse from the radiation of the other type (\al or \gm)is shown in light gray, while the vertical dashed line marks the beginning of the data interval used for the minimization.
  }
  \label{fig:avg_pulses}
 \end{figure*}

 Among well-established inorganic scintillators, cerium-doped yttrium aluminum garnet (\YAG) combines fast timing characteristics, good light yield, and exceptional mechanical and chemical stability.
 Its scintillation mechanism is governed by the allowed 5d--4f transitions of \ce{Ce^{3+}} ions \cite{VANEIJK1994546}, producing a broad emission band centered around $520-550$~nm, well matched to the spectral sensitivity of most silicon photodetectors.
 Early systematic studies demonstrated light yields of the order of $(1.0-2.5) \times 10^4$~photons MeV$^{-1}$ and a two-component decay time, i.\,e.\ a fast component of $\sim 70-90$~ns and a slow component of $\sim 250-300$~ns at room temperature, depending on the cerium concentration and excitation conditions~\cite{Moszynski:1994}.
 Recent developments in the engineering of \YAG have further enhanced its performance by reducing afterglow, accelerating scintillation decay, and improving light yield, where a value of $3.5 \times 10^4$~photons MeV$^{-1}$ has been reported~\cite{Zapadlik:2022}.
 A distinctive advantage of \YAG is its radiation hardness and non-hygroscopic nature, which enable stable operation in harsh environments.
 Irradiation studies with \gm rays and high-energy protons have shown negligible degradation of the scintillation performance up to doses and fluences relevant for high-energy physics experiments, confirming \YAG as a robust candidate for calorimetry and beam monitoring applications~\cite{Lucchini:2016}.
 Furthermore, the relatively fast rise time and the presence of distinct fast and slow scintillation components allow pulse-shape discrimination between different types of ionizing radiation~\cite{Ludziejewski:1997,Kim:2019crystal}.

 These properties have led to a widespread use of \YAG crystals in a broad range of applications, including X-ray and \gm-ray detection, charged particle detection, and beam diagnostics in particle accelerators.
 At the same time, \YAG crystals are also used in synchrotron-radiation facilities and electron microscopy as phosphor and scintillation screens, benefiting from their intense visible emission and mechanical robustness, while other applications include space instrumentation, radiation dosimetry, and industrial non-destructive testing, where long-term stability, non-hygroscopic behavior, and high radiation tolerance are critical requirements.
 
 In this work, we present a detailed experimental study of a \YAG scintillating crystal, focusing on its response to \gm and \al radiation.
 We quantify its light yield and quenching behavior.
 By employing a pulse reconstruction model, we extract the scintillation decay-time constants at different temperatures. In addition, we investigate the particle discrimination capabilities of \YAG using \al particles over a range of energies. These results are relevant for assessing the performance of \YAG in applications requiring precise energy reconstruction and detector response, including current and next-generation radiation detection systems.

\section{YAG:C\lowercase{e} scintillation}

 We performed a series of experimental measurements using a $10\times10\times10$~mm$^3$ \YAG crystal manufactured by Epic Crystal~\cite{EpicCrystal}, optically coupled to a silicon photomultiplier (SiPM) for scintillation-light readout.
 We used a Hamamatsu S13360-series Multi-Pixel Photon Counter with a $6\times6$ mm$^2$ active area, comprising $14,400$ microcells with a $50$-$\mu$m pitch~\cite{Hamamatsu_S13360}.
 To maximize light-collection efficiency, the four lateral faces of the crystal were wrapped with PTFE tape, while the face opposite the SiPM was left bare to allow direct line-of-sight for \al particles.
 The detector assembly was housed in an aluminum, vacuum-tight cylindrical chamber and connected to the electronics and DAQ systems.
 For each configuration of interest, digitized waveforms were collected and analyzed offline using custom Python routines to extract pulse parameters, apply event-selection criteria, and construct calibrated energy spectra.
 Further details on the experimental setup and signal readout are given in Ref.~\cite{Gironi:2025GAGG}.

\subsection{Scintillation signal reconstruction}
\label{sec:sim}

 In order to extract the scintillation parameters, we reproduced the shape of particle-induced physics pulses from both \gm and \al interactions starting from individual SiPM single-cell pulses.
 Measurements with \al particles were performed using a \Am source placed inside the chamber at distances of approximately $5$ and $10$~cm from the crystal, while a \Na source positioned outside the chamber was used for \gm rays; the former provides an \al peak at $5.49$~MeV (B.\,R.\ $\sim 85\%$) when a vacuum is established inside the chamber and a \gm peak at $59.5$~keV (B.\,R.\ $\sim 36\%$) regardless of the pressure.
 The \Na source provides a peak at $511$~keV. 

 The single-cell pulses were recorded and isolated during data-taking runs using a low trigger threshold. From this dataset, we constructed physics-like scintillation signals by summing a large number of single-cell pulses whose occurrence times were randomly generated according to a double-exponential distribution characterized by a short ($\tau_{\text{short}}$) and a long ($\tau_{\text{long}}$) time constant, as described in detail in Ref.~\cite{Gironi:2025GAGG}. The relative weight $F$ defines the ratio between the two components.

 The scintillation parameters ($\tau_{\text{short}}$, $\tau_{\text{long}}$, and $F$) were extracted by minimizing the difference between the experimental scintillation pulses and the reconstructed ones, considering only the final part of the rising edge and the decay of the pulse.
 In particular, we used the sum of squared differences calculated point-by-point between the experimental scintillation pulses and the reconstructed ones as the figure of merit to be minimized, equivalent to a non-normalized $\chi^2$ under the assumption of uniform uncertainties.
 In the minimization procedure, we deliberately excluded the initial portion of the rising edge, as it is strongly affected by signal filtering and triggering/alignment threshold effects, and instead used an effective double-exponential description of the pulse evolution by focusing on the decay region.
 Therefore, this model aims to provide a phenomenological description of the pulse shape in the decay region, rather than a complete modeling of the signal formation over the entire time range.
 Despite this, as shown in Fig.~\ref{fig:avg_pulses}, the reconstruction obtained with the optimized parameters also reproduces the full pulse shape with good accuracy.
 
 The results obtained from the reconstruction of physics-like scintillation signals are summarized in Table \ref{tab:sim_results}.
 It is worth noting that only in a few cases in the literature an additional, slower scintillation component, on the order of 1 $\mu s$, is taken into account~\cite{Kim:2019crystal}.
 In our modeling, we chose to neglect this contribution because it is marginal and would unnecessarily complicate the parameterization by introducing two extra degrees of freedom (the scintillation time constant and the corresponding fractional yield).
 Including this component would reduce the stability of the fit without providing any significant improvement in the description of the data.

 This method also allows us to determine the number of optical photons contributing to a single scintillation pulse by estimating the number of single-cell events required to reconstruct the acquired signals.
 We extract the number of fired cells from the ratio of the integrated pulse area to that of the single-cell template.
 This value is then converted into the number of incident photons by correcting for the deposited energy, the photon detection efficiency, and geometrical losses via an approximation suggested by the SiPM manufacturer~\cite{HPK:web_formula}.
 Applying this procedure to \gm events enabled the evaluation of the light yield, which was found to be around 19,000~photons MeV$^{-1}$.
 We do not account for optical transport effects inside the crystal, such as self-absorption, given its small dimensions.
 However, studies on garnets similar to \YAG\ have shown that the impact on cm-scale crystals is limited~\cite{Uchida:2021CeGAGG}.
 Moreover, by performing measurements at different pressures with the \Am source and evaluating the \al-particle energy using Geant4 Monte Carlo simulations~\cite{GEANT4:2002zbu}, it was possible to investigate the quenching as a function of the energy.

 \begin{table}[tb]
  \centering
  \caption{Scintillations parameters for \YAG extracted from the reconstruction of physics pulses from single-cell templates.}
  \vspace{5pt}
  \setlength{\tabcolsep}{8pt}
  \begin{tabular}{lrrr}
   \hline \\[-8pt]
   Parameter                   &\al       &\gm            \\[2pt]  
   \hline   \\[-7pt]
   $\tau_\mathrm{short}$ [ns]  &$ 63$    &$ 67$           \\[5pt]
   $\tau_\mathrm{long}$ [ns]   &$273$    &$245$           \\[5pt]
   $F_\mathrm{short}$          &$0.10$   &$0.15$          \\[5pt]
   \hline
  \label{tab:sim_results}
  \end{tabular}
 \end{table}

\subsection{Quenching of alpha particles}

 The different light yield in scintillators for \al --and, more generally, highly ionizing-- particles compared to electrons and \gm rays per unit deposited energy arises from the high density of charge carriers along the particle track, leading to saturation of the luminescence centers~\cite{Lecoq:2020itu}.
 It is typically parameterized in terms of a Quenching Factor (\QF), defined as the ratio of the light yield for \al particles or ions to that for electrons at equal deposited energy.
 The \QF is energy dependent and reaches a minimum where the stopping power is maximal; in many crystalline scintillators, this occurs around 1 MeV for \al particles~\cite{Tretyak:2009sr}.

 By tuning the pressure inside the vacuum chamber, we performed measurements with \al particles depositing in the scintillator a variable fraction of the available energy from the \Am decay, from $0$ up to the full $5.49$ MeV.
 We acquired data at two source–crystal distances, $49$ and $97$ mm, and with different gases: air, argon, and helium.
 The pressure was varied from $\sim 1$ mbar up to several hundred mbar; the maximum value depended on the gas and the distance and was set by the condition that the \al particles still reached the crystal.
 At low pressure, the \al particles retain nearly their full decay energy, whereas increasing the pressure causes energy loss in the gas before they enter the crystal.

 To quantify the energy lost by the \al particles in the gas along the path from the source to the crystal and the energy deposited in the crystal itself, we performed Monte Carlo simulations of the various configurations using the Geant4 toolkit.
 From each simulated spectrum, we extracted the mean energy deposited in the detector.
 To evaluate the light yield for \bt/\gm events, we relied on the $511$-keV peak from \Na for linear calibration, acquiring a few runs with the \Na source over the measurement campaign.
 Owing to the low ionization density of $\beta/\gamma$ interactions, the light yield in the MeV range can be considered approximately constant to first order.
 Deviations from linearity could occur below about $100$ keV~\cite{Beretta:2017bft}, which is why we did not use the $59.5$-keV peak from \Am. 
 From the fit, we estimated the light yield expected at the energies corresponding to \al-particle energies at the different pressures.
 The resulting ratio was the \QF.

\begin{figure}[tb]
  \centering
  \includegraphics[width=1.\columnwidth]{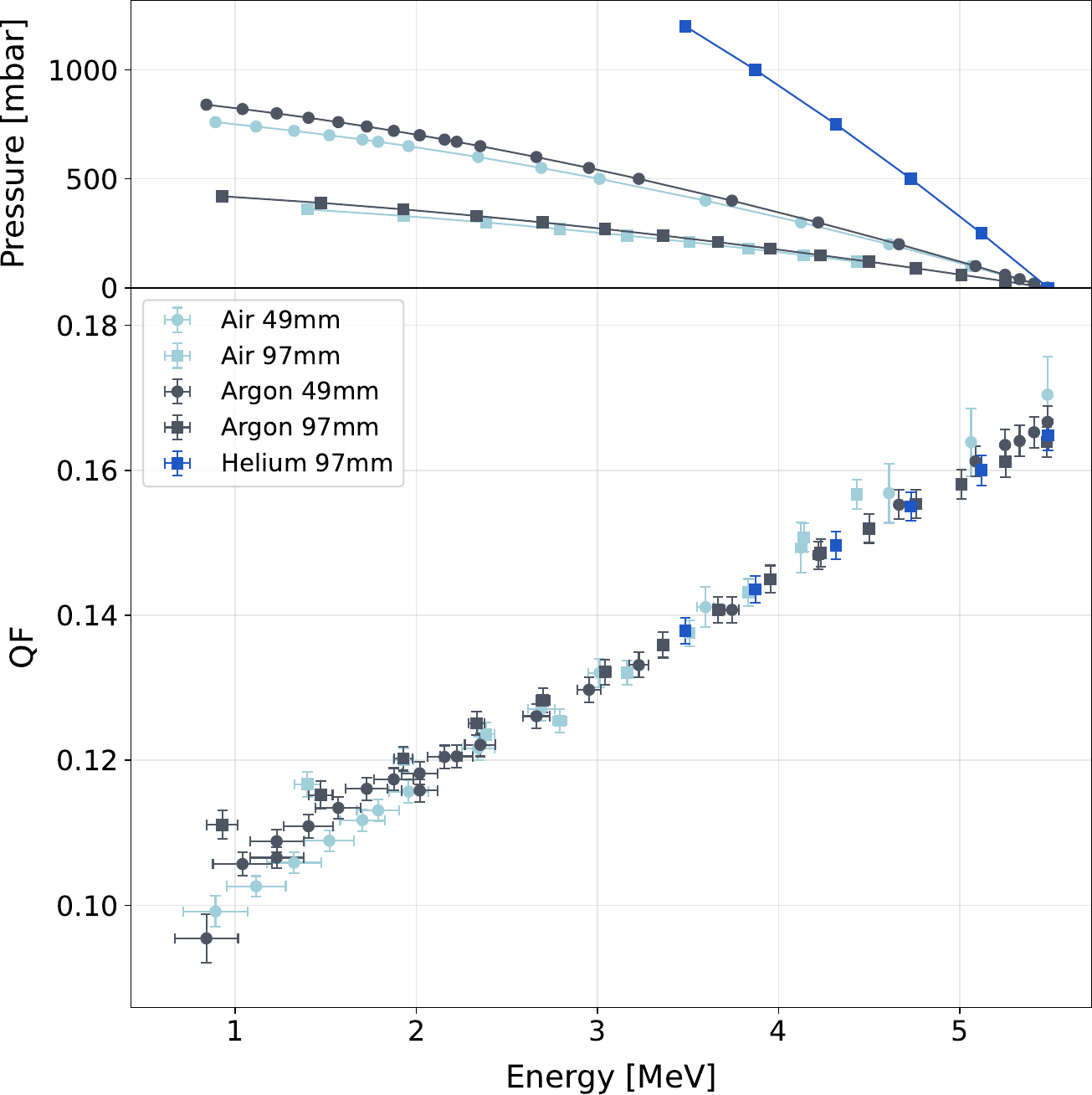}
  \caption{Pressure {\it (top)} and \QF of \YAG~{\it (bottom)} as a function of \al energies.
   Different series were acquired at fixed source-detector distances of $49$ and $97$ mm and with different gases: air, argon, and helium.
   We considered a systematic uncertainty corresponding to a $\pm 1$ mm variation in the source-detector distance, which defines the error bars on the energy value, and propagated the uncertainty from the linear \bt/\gm calibration as the error on \QF.}
  \label{fig:QF}
 \end{figure}

 Figure~\ref{fig:QF} shows the results for the different gas and distance configurations.
 The upper panel illustrates how the energy loss in the gas varies significantly among different gases and is affected by the source-detector distance.
 This behaviour can be understood by considering together the gases' masses and ionization properties.
 In particular, although argon has a higher atomic number than air, its energy loss is smaller because of its higher mean excitation energy, which affects the stopping power according to the Bethe-Bloch formula~\cite{Qadr:2017airar}, while helium exhibits a much lower stopping power due to its low atomic number and relatively high mean excitation energy.
 The actual values of \QF are presented in the bottom panel, ranging from about $0.17$ down to $0.10$ as the \al energy decreases from $5.49$ to 1 MeV.
 We included error bars on the computed values, which were obtained by propagating the uncertainty on the energy.
 For the energy estimate, due to limited precision in determining the source-detector distance, we repeated the simulation with its value varied by $\pm 1$~mm, and we took the resulting differences in the simulated $\alpha$ energies as the corresponding uncertainty; in addition, we propagated the uncertainty from the linear \bt/\gm\ calibration.

 In our measurements, the determination of \QF is limited to energies above approximately $1$ MeV.
 This constraint arises from the presence of the $59.5$-keV \gm line of \Am, which prevents reliable reconstruction of the \al peak, when this is shifted towards very low energies.
 As a possible way to probe the sub-MeV region in future studies, one could reduce the thickness of the \YAG crystal in order to decrease the ratio between the two contributions in favor of the latter.
 
\subsection{Scintillation at different temperatures}

 The temperature dependence of the scintillation response of \YAG was investigated by performing measurements inside a controlled climate chamber over the range from room temperature at $20^\circ$ C down to $-50^\circ$ C (approximately $295$–$225$ K).
 The $59.5$-keV peak from \Am was used as a reference, and the spectrum was calibrated at each temperature to the single-cell response of the SiPM.
 Corrections were applied to account for the temperature dependence of the photon detection efficiency through its dependence on the overvoltage, which in turn varies with temperature.
 The impact of the SiPM optical cross-talk probability was neglected, as it remained at the level of a few percent and thus below our experimental sensitivity~\cite{Hamamatsu_S13360}.

\begin{figure}[tb]
  \centering
  \includegraphics[width=1.\columnwidth]{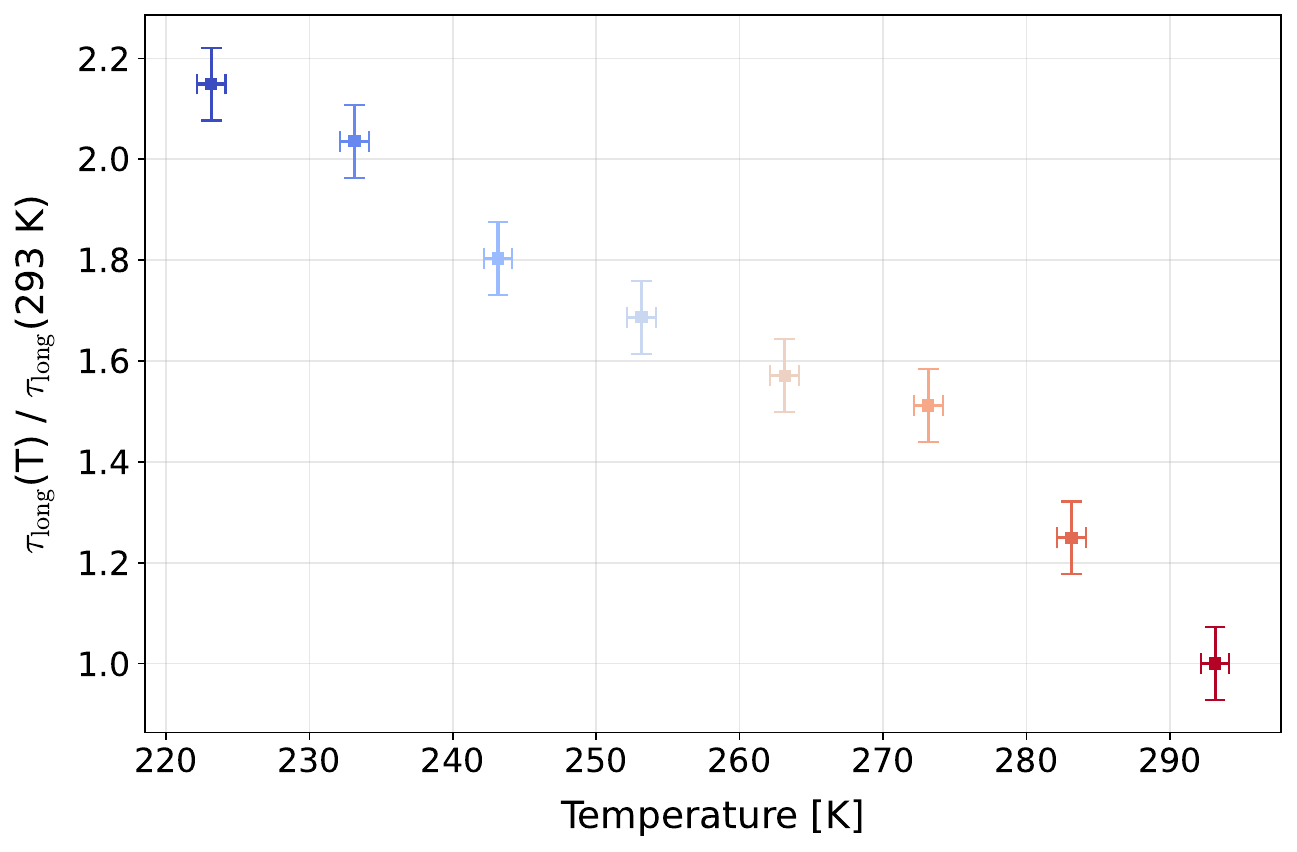}
  \caption{Evolution of $\tau_\mathrm{long}$ of \YAG as a function of temperature normalized to the value at room temperature.
   Over the investigated range (approximately $225-295$~K), $\tau_\mathrm{long}$ increases by approximately a factor of two.}
  \label{fig:CC}
 \end{figure}

 No significant variation in the scintillation light output was observed with decreasing temperature; however, a change in the time evolution of the scintillation pulse was recorded.
 Possible temperature-induced changes in the SiPM single-cell response are intrinsically accounted for in the reconstruction procedure, since the scintillation pulses are generated using single-cell waveforms measured at the corresponding temperature. As a consequence, variations in the SiPM pulse shape do not bias the extraction of the scintillation decay constants.
 The variation of the scintillation time is illustrated in Fig.~\ref{fig:CC}, where the ratio between $\tau_\mathrm{long}$ at room temperature and at the selected temperature is shown as a function of temperature.
 The long decay-time component was extracted at each temperature by means of the pulse-reconstruction simulation described in Sec.~\ref{sec:sim}.
 It exhibits an increase with decreasing temperature, indicating a progressive slowdown of the emission kinetics.
 Over the investigated temperature range, its value increases by approximately a factor of two.

 The short decay-time component remained overall constant within the experimental uncertainty.
 This parameter is more sensitive to factors such as the selection criteria for the single-cell pulse, the choice of the minimization seed, and the definition of the start and end points of the fitting range used in the minimization procedure.
 Therefore, no significant variation of $\tau_\mathrm{short}$ could be observed.

\section{Pulse-shape analysis}
\label{sec:analysis}

 Due to the scintillation kinetics in \YAG, which depends on the interacting radiation, \al- and \gm-induced events produce pulses with distinct temporal profiles.
 As illustrated in Fig.~\ref{fig:avg_pulses}, the characteristic decay differs for the two interaction channels, resulting in measurable variations in the waveform evolution.
 These differences enable the application of pulse-shape discrimination techniques to classify events.

 To this end, we tested different strategies to quantify the separation between the two pulse families.
 We implemented a procedure based on a distance metric, comparing each normalized waveform point-by-point with reference \al and \gm, constructed by averaging events associated with the corresponding calibration peaks~\cite{Gironi:2025GAGG}.
 We also explored the efficacy of discrimination parameters based on pulse timing.
 We used a {\it cumulative-pulse} parameter, defined as the time index at which the cumulative integral of the waveform reaches a given fraction of its maximum, and a {\it partial-charge} parameter, obtained by integrating the pulse over a selected time window within the decay-time region.
 The three parameters were found to be highly correlated and showed comparable discriminatory performance; we selected the partial-charge parameter, as it demonstrated slightly superior sensitivity.
 The discrimination capability of this parameter is further improved by smoothing the normalized pulses to mitigate the effect of dark counts.
 
 The time window used to compute the partial-charge parameter is defined to maximize the discrimination power, extending from the point where the two pulses diverge most, around $400$~ns after the pulse maximum, to approximately the end of the pulses.
 A clustering algorithm (implemented via the \textit{Scikit-learn} library) is then used to assign each waveform to the closest reference.
 Defining the \textit{Discrimination Power} as $D \equiv \left| \mu_{\al} - \mu_{\gm} \right|/\sqrt{\sigma_{\al}^2 + \sigma_{\gm}^2}$, we obtain a separation of $D \approx 2.3$ between the $\alpha$ and $\gamma$ distributions when the full \al energy is deposited in the crystal, and $D \approx 1.3$ when \al events produce signal amplitudes comparable to those of the $511$-keV peak.
 The discrimination accuracy, evaluated on a control sample, is at least $93\%$ at the \al full-energy peak.
 At lower energies, the accuracy degrades—reaching approximately $70\%$ at the $511$-keV peak, as expected from the lower signal-to-noise ratio.

\begin{figure}
  \includegraphics[width=1.\columnwidth]{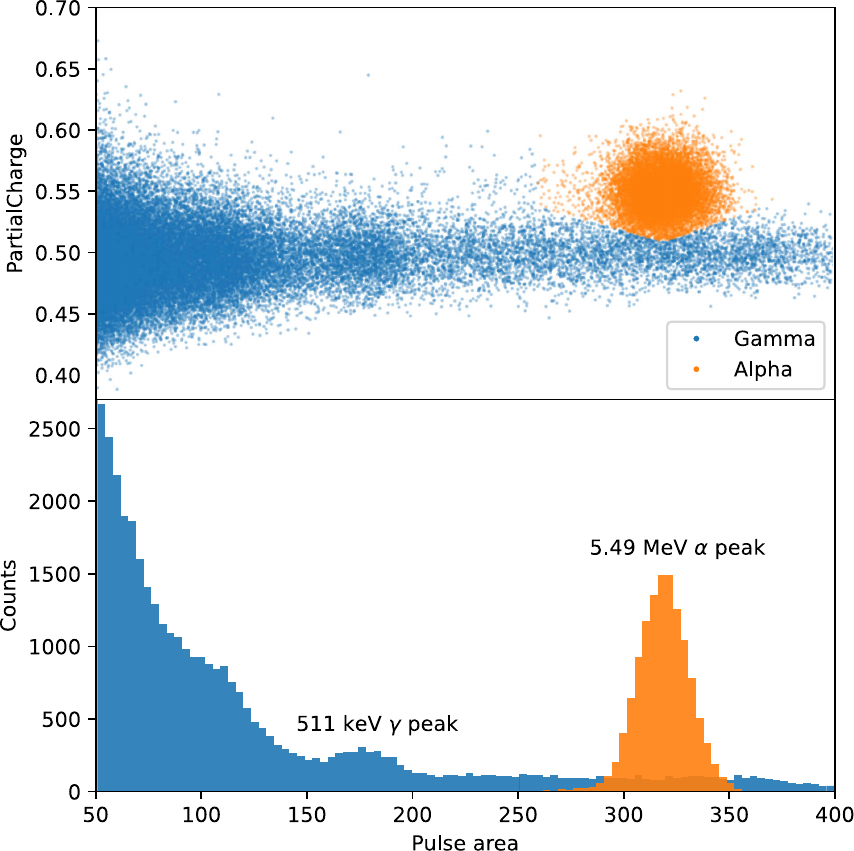}
  \caption{Discrimination of \al and \gm events from the $5.49$-MeV and $511$-keV peaks during a run with both \Am and \Na sources at a pressure aligning the \al peak with the \gm region.
   {\it (Top)} Partial-charge parameter {\it vs.} pulse total integrated area (collected charge), showing two populations identified by clustering. At this energy of the \al,  $D \approx 2.3$.
   {\it (Bottom)} Energy spectrum with the isolated populations.}
  \label{fig:PSD_0mbar}
 \end{figure}

 The discrimination capabilities are illustrated in Fig.~\ref{fig:PSD_0mbar}, referring to a run in which both \Am and \Na sources were present.
 The upper panel shows the distribution of the partial-charge parameter for events within the same energy range as the lower panel, which displays the corresponding energy spectrum.
 A clear separation between particle types is achieved for deposited energies above $0.5$~MeV, allowing events to be reliably attributed to \al or \bt/\gm interactions through pulse-shape analysis.

\subsection{Summary \& outlook}

 In this work, we presented a detailed experimental investigation of the scintillation properties of a \YAG crystal and its response to \al and \gm radiation.
 By reconstructing physics pulses from single-cell SiPM events, we extracted the fast and slow scintillation components, which were found to be approximately $70$ and $250$~ns, respectively. We verified that the two-component model reproduces the overall pulse evolution for both \al and \gm events by adjusting the relative weights of the two components.
 
 The use of a vacuum-tight chamber enabled probing a wide range of \al energies by varying the pressure and allowed determination of the energy-dependent \QF for \al particles over the range $\sim 6$ to $\sim 1$ MeV, which decreases from about $0.17$ to $0.10$ with decreasing energy. 
 This setup also allows future extension of the study into the sub-MeV region by optimizing the shape of the \YAG crystal.

 We studied the temperature dependence of the scintillation process between $225$ and $295$~K and observed approximately a factor-of-two increase in the long decay constant, while the short component remained largely unchanged.

 Finally, we tested different pulse-shape discrimination methods to assess the discrimination capabilities of \YAG.
 Clear separation between \al and \gm events depositing the same energy was achieved by employing a partial-charge parameter, allowing for extraction of the pulse shape information.

 Overall, this study provides a comprehensive characterization of \YAG scintillation performance and establishes a solid foundation for its use in next-generation radiation-detection systems requiring stable response, particle identification, and reliable operation under diverse environmental conditions.
 
\acknowledgments
 The authors acknowledge the mechanical workshop team of INFN Milano-Bicocca for their constant and constructive support in the design and construction of the setup, and Luca Brambilla, Elena Pedrazzoli, and Giorgio Lauriola for their assistance with the measurements.
 This work is performed in the framework of the European Pathfinder Open project UNICORN (GA 101098649) and has been partially supported by the Italian Ministry of University and Research (MUR) through the grant Progetti di ricerca di Rilevante Interesse Nazionale (PRIN grant no.\ 2020H5L338).

\bibliography{ref} 

\end{document}